\documentclass[sigconf]{acmart}
\usepackage[utf8]{inputenc}
\usepackage{xspace}
\usepackage{booktabs}
\usepackage{mathtools}
\usepackage{cleveref}
\usepackage{xcolor}
\usepackage{amssymb}
\usepackage{breqn}
\DeclarePairedDelimiterX\Basic[1](){ #1}

\newcommand{\MakeEst}[1]{\expandafter\hat#1}

\newcommand{\exleak}{EL\xspace}
\newcommand{\statleak}{SL\xspace}

\setcopyright{acmcopyright}
\setcopyright{acmlicensed}
\setcopyright{rightsretained}
\acmYear{2018}
\copyrightyear{2018}
\setcopyright{acmlicensed}
\acmConference[WiSec '18]{WiSec '18: Proceedings of the 11th ACM Conference on Security and Privacy in Wireless \& Mobile Networks}{June 18--20, 2018}{Stockholm, Sweden}
\acmBooktitle{WiSec '18: Proceedings of the 11th ACM Conference on Security \& Privacy in Wireless and Mobile Networks, June 18--20, 2018, Stockholm, Sweden}
\acmPrice{15.00}
\acmDOI{10.1145/3212480.3212508}
\acmISBN{978-1-4503-5731-9/18/06}

\graphicspath{{./figures/}}

\begin{document}

\title{I Send, Therefore I Leak: Information Leakage in Low-Power Wide Area Networks}

\author{Patrick Leu}
\affiliation{%
  \institution{ETH Zurich, Switzerland}
}
\email{patrick.leu@inf.ethz.ch}

\author{Ivan Puddu}
\affiliation{%
  \institution{ETH Zurich, Switzerland}
}
\email{ivan.puddu@inf.ethz.ch}

\author{Aanjhan Ranganathan}
\affiliation{%
  \institution{Northeastern University, USA}
}
\email{aanjhan@northeastern.edu}

\author{Srdjan \v{C}apkun}
\affiliation{%
  \institution{ETH Zurich, Switzerland}
}
\email{capkuns@inf.ethz.ch}

\begin{abstract}
{Low-power wide area networks (LPWANs), such as LoRa, are fast emerging as the preferred networking technology for large-scale Internet of Things deployments (e.g., smart cities). Due to long communication range and ultra low power consumption, LPWAN-enabled sensors are today being deployed in a variety of application scenarios where sensitive information is wirelessly transmitted. In this work, we study the privacy guarantees of LPWANs, in particular LoRa. We show that, although the event-based duty cycling of radio communication, i.e., transmission of radio signals only when an event occurs, saves power, it inherently leaks information. This information leakage is independent of the implemented crypto primitives. We identify two types of information leakage and show that it is hard to completely prevent leakage without incurring significant additional communication and computation costs.}
\end{abstract}

%\keywords{Internet of Things, LoRaWAN, Information Leakage}

\maketitle
%!TEX root =  lora-privacy-main.tex
\section{Introduction}
The recent advancements in communication and computing technologies have led to the rapid proliferation of connected devices (Internet of Things).
It is expected that by the year 2020, billions of \emph{things} capable of sensing and wirelessly communicating data will be deployed, effectively creating numerous smart ecosystems such as smart homes, cities, and industries. The choice of wireless technology is critical as it directly impacts the network's power consumption, coverage area, achievable data rate, and deployment cost. Popular wireless connectivity standards, such as WiFi and Bluetooth, offer high data throughput and optimal power consumption but are limited in the communication range. Cellular communication technologies such as GSM and LTE have been specifically designed for high data rate applications and can operate over several kilometers. However, these technologies are not optimized for power consumption and devices using them require frequent recharging or power source replacement. Smart ecosystems such as smart cities and smart industries rarely require high data rate. But large coverage areas, spanning a few hundred meters to few kilometers, as well as ultra-low power consumption are imperative for large-scale deployments.

Low-power wide area networks (LPWANs), such as LoRa~\cite{lora}, SigFox~\cite{sigfox}, NB-IOT~\cite{ratasuk2016nb}, and Weightless~\cite{weightless} are fast emerging as the preferred wireless technology for implementing large-scale, smart ecosystems where data rate is less important than the communication range, battery life or deployment cost.
LPWANs are specifically designed to enable low-data-rate (typically a few kbps), long-range communications (up to tens of kilometers) on battery-operated \emph{things} (also called end devices). Today, there exist already numerous LPWAN-enabled end devices, such as sensors that detect garbage levels in trash bins~\cite{smartbin} and automatically notify the waste removal trucks, equipment maintenance and failure sensors that notify managers of status, occupancy sensors~\cite{ascoel} that detect presence or absence of people in a space, push button sensors~\cite{ascoel} that can be used to initiate maintenance calls, pest detection sensors~\cite{pest}, and many more.  Many of these sensors are deployed in application scenarios where sensitive information is broadcasted. Therefore, it is important to analyze and understand the security and privacy guarantees of LPWANs.

In this work, we focus on the privacy guarantees of LPWANs, in particular LoRa.
One of the key characteristics of LPWAN technologies like LoRa is ultra-low power consumption. Typically, LoRa-enabled sensors only turn on their radio when an event has been detected and there is an immediate need to communicate the occurrence of the event. During the remainder of the time, the end devices fully turn off their communication hardware to save power. In this work, we show that such an event-driven communication leaks information, independently of the implemented cryptographic primitives. For example, a parking space sensor that transmits wireless LoRa packets whenever a car pulls over already leaks information regarding the "presence of car" event, irrespective of the crypto primitive deployed. Although such leakage can happen in other networks, due to aggressive duty cycling and large communication range (several kilometers), LoRa networks are particularly vulnerable to such attacks. An attacker that strategically places a few devices across the city will be able to collect a number of business and private data.

This is in contrast to communication networks like cellular and WiFi, where the mere communication typically does not reveal sensitive information (i.e., communication is not event-driven) or to sensor networks, which have limited communication range and therefore require the attacker to get physically close to the sensors. In terms of leakage, LPWANs combine the worst - long communication range and event-driven communication. Therefore, it is fairly trivial for an attacker to deduce occurrences of an event or a set of events, equipped with just a simple receiver (e.g., a software defined radio~\cite{ettus}). We note that although our analysis largely focusses on LoRa (due to its popularity), the conclusions of this work are generic and fundamental to all existing low-power wide area networking technologies.

Several recent works on LoRa and other low-power wide area networks have focussed on understanding the performance guarantees, such as scalability~\cite{bor2016lora,georgiou2017low,adelantado2017understanding} and channel capacity~\cite{adelantado2017understanding,mikhaylov2016analysis} of the network. Some others focussed on studying the impact of physical layer settings on the data rate and energy efficiency of the communication system~\cite{ochoa2017evaluating,casals2017modeling}. Several vulnerabilities, such as replay attacks~\cite{tomasin2017security}, acknowledgement spoofing~\cite{yang2017lorawan}, physical key extraction~\cite{aras2017exploring} and device fingerprinting~\cite{Robyns:2017:PFL:3098243.3098267} have already been demonstrated on LoRaWAN's security protocols and their implementations. To the best of our knowledge, the privacy implications of LoRa-like low-power wide area networks have not been extensively studied so far.

Specifically, in this work, we make the following contributions. We show that LoRa-like wireless networking technologies inherently leak information due to an event-based communication strategy.
We identify two types of information leakage: (i) existential leakage and (ii) statistical leakage.
We show that existential leakage can be, to an extent, prevented by adding dummy transmissions. However, statistical leakage is hard to obfuscate without incurring additional communication cost and affecting scalability, due to the increased usage of the communication channel.
By means of simulations, we show that it is challenging to achieve full prevention of information leakage without compromising on the ultra-low-power guarantees of the system. For an obfuscator whose knowledge is limited, our results indicate that optimal obfuscation may not be achievable even for rare anomalies, under a power constraint that limits the rate of obfuscation packets to that of real messages. We also quantify the attacker's performance under this power constraint and compare it to the case of optimal obfuscation. Thus, through this work, we highlight the tension between designing low-power wide area networks and achieving foolproof privacy guarantees.

%!TEX root =  lora-privacy-main.tex

\section{Low-power Wide Area Networks}
\label{sec:background}
Low-power wide area networks are designed with the key objectives of long distance communication (wide area coverage), ultra low power consumption at the end devices and low deployment cost. LPWANs achieve these design objectives by leveraging the low data rate requirements of the majority of IoT applications. In this section, we give an overview of the network architecture, the communication protocols and the security properties of LPWANs with a specific focus on LoRa.

\subsection{Network Architecture}
\begin{figure}[t]
	\begin{center}
		\includegraphics[width=0.75\columnwidth]{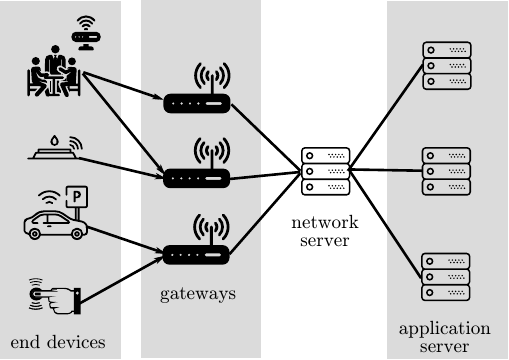}
		\caption{Network architecture: Application-specific end nodes reach their network server via a gateway in proximity. From there, the data is routed to the respective application server.}
		\label{fig:network}
	\end{center}
\end{figure}

LPWANs are implemented using a star-of-stars network topology as shown in Figure~\ref{fig:network}. End devices are application specific (e.g., parking lot occupancy sensors, motion sensors) and are connected to one or many \emph{gateways}. Gateways act as transparent relays between end devices and a network server. One of the key differences between LPWAN and conventional cellular networks is that end devices are not required to associate to a specific gateway but are only associated with the network server. This makes it possible for end devices to communicate with more than one gateway at the same time. Furthermore, end devices can be mobile and connected to the network without any complex signalling and handoff mechanisms. Gateways use cellular or ethernet as backhaul to connect to the network server, which then forwards the information to corresponding application servers for processing. It is the responsibility of the network server to filter redundant messages forwarded by multiple gateways, perform security checks and, if necessary, schedule acknowledgements, therefore reducing the complexity of end devices and gateways.

\subsection{Communications System}
One of the key objectives of LPWANs is to enable low-power operation of end devices. Therefore, LPWANs aggressively duty cycle end devices by turning on their radio only when there is an event and there is a need to communicate its occurrence. During the remainder of the time, the end devices are fully turned off to save power. In case the gateway needs to communicate with the end device, it can only do so during an a-priori agreed time schedule. For example, in LoRa, the end devices can be operated in three different modes: Class A, B and C. Class A mode is a mandatory mode where the end devices open two receiving slots immediately after an event-triggered transmission. The receiving slots \emph{may be} used to get a response or acknowledgement back from the gateway or the application server. The optional class B and C modes allow the end devices to receive data more frequently from the gateways and are intended for applications without any power constraints. In this work, we mainly focus on the mandatory class A operation mode, however the results are in general applicable to the other operating modes as well. The end devices do not execute any form of channel sensing or signalling prior to transmission. As soon as there is an event, the end device instantly broadcasts a message to communicate the occurrence of the event. In other words, unlike in a majority of wireless networks, there are no complex medium access protocols minimizing the control signalling overhead. Furthermore, the long communication range allows a one-hope network topology. This reduces the complexity at both the end devices and the gateways, as they no longer require complex listen and forward mechanisms that are commonly implemented in multi-hop wireless sensor networks. The long communication range is achieved by choosing physical-layer techniques that transmits more energy per each data symbol while compromising on data rate. For example, LoRa uses a modulation technique based on chirp spread spectrum~\cite{berni1973utility} with forward error correction that enables the end devices to communicate over long distances. The use of sub-1-GHz frequencies also results in less attenuation and better signal penetration through walls and other environmental obstacles.

\subsection{Security and Privacy}\label{sec:lora_privacy}
Most LPWAN technologies opt for symmetric key cryptography to provide support for end device message authentication and application payload encryption. For instance, in LoRa, each end device is equipped with a unique 128-bit secret key (\emph{AppKey}), which is used to derive two session keys; one each for sharing with the network server or the service provider (\emph{NwSKey}) and the application server (\emph{AppSKey}). The application payload can be encrypted using the \emph{AppSKey}, the \emph{NwSKey} can be used to generate the message integrity code. This prevents man-in-the-middle scenarios where the network or service provider acts malicious and wants to eavesdrop on all the traffic between end devices and the application server. However, in practice it has been observed that end devices typically do not encrypt the application payload and the network service provider can eavesdrop on the communication between end devices and the application server. The communication links between the gateway devices and the network server themselves are secured using standard TLS/IPSec.

%========================================
%========================================

\subsection{Applications}
\label{sec:applications}
Smart cities, smart homes and buildings as well as industrial IoT applications are prime examples of segments that will greatly benefit from LPWAN technologies.\\

\noindent\textbf{Industrial applications:} A typical industrial building may contain a reception or welcome center, meeting rooms, parking area for employees and visitors, cafeteria, manufacturing floor, lobby or waiting areas for visitors etc. Parking spots\footnote{\url{https://www.pnicorp.com/placepod/}} can be equipped with end devices or sensors that detect the presence of a vehicle and transmit information using LPWAN to the central servers. Occupancy or vacancy sensors\footnote{\url{https://www.thethingsnetwork.org/community/thatcham/post/occupancy-sensor}} can be installed in meeting rooms and manufacturing floors to remotely control the heating, ventilation and air-conditioning systems. Furthermore, vacancy sensors can be used to communicate unused discussion rooms. Individual entry and exit of personnel, manufactured products, raw materials can also be tracked using access sensors. Push button sensors enable user interaction with the network and they can be used to indicate an equipment failure, accidents or for simple command and control applications.\\

\noindent\textbf{Smart homes and cities:} LPWAN technology can help accelerate IoT adoption into several urban and home applications. Today, there is already a vast variety of commercially available sensors that can seamlessly integrate into the LPWAN infrastructure such as burglar alarm and intrusion sensors, emergency care, garage door entry systems, traffic management, rodent trap sensors, water and chemical leakage sensors.

To summarize, LPWAN communications are event-driven, application-specific and have long communication range. They are power efficient, which makes them attractive to a wide variety of privacy sensitive business and personal applications. In this paper, we show how the properties of LPWANs leaks information independent of the implemented cryptographic primitives.

%!TEX root =  lora-privacy-main.tex
\section{Information Leakage in LPWAN}
\label{sec:privacy_implications}
%========================================
\begin{figure}[t]
	\begin{center}
		\includegraphics[width=0.75\columnwidth]{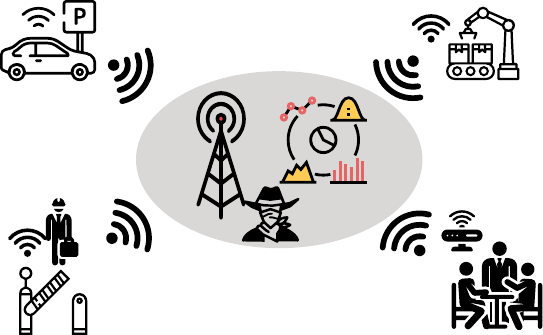}
		\caption{An end device transmits a message to the application server upon sensing an event. A passive adversary will be able to deduce sensitive information such as arrival and departure times of employees, the number of visitors, meeting times etc.}
		\label{fig:leakage-scenario}
	\end{center}
\end{figure}
%========================================

We motivate the problem of information leakage in low power wide area networks with the following example scenario. As illustrated in Figure~\ref{fig:leakage-scenario}, a smart parking sensor will transmit a packet as soon as a vehicle occupies or leaves the parking space. Similarly, any movement or presence of personnel in an office space will trigger radio transmissions in occupancy sensors. Every access attempt of employees to their workspaces will result in radio communication. The long communication range of LPWANs allows these messages to be received several hundred meters or even kilometers away. In other words, an attacker located several hundred meters away will be able to infer sensitive information such as arrival and departure times of employees, meeting schedules, presence or absence of personnel within a building, etc.

In this work, we assume  a passive adversary equipped with one or more receivers (e.g., built using a software defined radio) collecting end device transmissions. We assume a LoRa network architecture where a set of end devices (EDs) are within communication range of at least one gateway. Note that in LoRa the end devices are not associated with one specific gateway. Any gateway within communication range can receive the an end device's signals. The gateway forwards the message to the network server for further processing. We focus on real-time applications, as described in Section~\ref{sec:applications}, in which the EDs immediately transmits a message to the application server as soon as it senses an event. We assume that the attacker has a priori knowledge regarding the application associated with an end device transmission. It has already been shown that multiple LoRa end devices can be uniquely identified using physical layer fingerprinting techniques~\cite{Robyns:2017:PFL:3098243.3098267}. Furthermore, the packet header information is transmitted in the clear. This makes it trivial for an eavsedropper to associate signal transmissions to specific EDs. Note that LPWAN service providers need to bill their costumers, and therefore have a mapping between EDs' MAC addresses and end users. Even less privileged attackers can still narrow down the origin of messages, e.g. by means of multilateration.

To the best of our knowledge, no LPWAN-enabled end devices currently implement any form of obfuscation or privacy enhancing technique. Therefore, today it is trivial for a passive adversary to deduce event properties. For example, in the case of a meeting room occupancy sensor, an attacker can deduce the meeting times and frequency by simply observing the LoRa packet transmissions. Note that we focus on time sensitive (i.e., delay intolerant) applications. Therefore, aggregation and periodic transmission techniques are no viable options.

Given this constraint and the network topology of LoRa, particularly not allowing direct sensor-to-sensor communication, the only strategy for EDs to obfuscate their traffic is to add dummy packets to their transmissions. Dummy packets are simply packets containing random data (of the same length as a typical communication). Since each LoRa packet payload is encrypted (cf.\ Section \ref{sec:lora_privacy}), dummy packets are indistinguishable by the attacker from packets containing meaningful information, but can be filtered out by the application server. The obfuscation mechanism, i.e., the decision when to transmit dummy messages, is important as it directly impacts cost. Firstly, each additional packet transmitted consumes energy, reducing the lifetime of the ED. Secondly, to minimize ED complexity, LPWANs allow EDs to access the communication channel in a random and uncoordinated fashion without any complex medium access controls. Therefore, each additional dummy packet transmitted also reduces channel availability. Finally, LPWAN service providers typically charge for every additional transmission above a certain threshold agreed upon a-priori.

In the remaining sections, we analyze which kind of information an attacker is able to obtain by observing LoRa traffic and to which degree she can still infer information, even when an obfuscation mechanism is in place.

\subsection{Existential Leakage (\exleak)}\label{sec:existance_leak}
\emph{Existential Leakage} (\exleak) covers all cases where transmission of a single message implies the occurrence of a real-world event. For instance, when using a presence sensor in an industrial application (cf. Section \ref{sec:applications}), existential leakage happens whenever a sensor transmits a message because of a vehicle occupying a particular parking space, whenever a sensor transmits a message because of a meeting room becoming empty, whenever a message is sent because a push button is pressed to indicate an equipment failure, to name a few.
This type of transmission behaviour is a problem particularly in LPWAN sensors because, given the constraints of the technology, having very simple sensors dedicated only to one function, that transmit only if a particular event occurs, is the norm. Whereas in many other wireless technologies there might be either multiple or no real-world events underlying any single message transmission, in LPWANs, the event space is usually very small, most times binary (e.g. parking lot is empty or occupied), and transmissions usually only occur on state transitions (e.g. parking lot is now occupied).

The goal of the attacker exploiting \exleak is to detect the presence of real messages. If no obfuscation is in place, this is trivial and can be done by simply eavesdropping on the LPWAN channel. An attacker that recognises if at least a real message is sent in a time interval of her choosing, managed to exploit \exleak. For instance, an attacker is exploiting \exleak if she detected that a rodent sensor sent a ``mouse captured'' message (as opposed to e.g. only empty messages) on a particular day.

It is important to understand the difference between an event and a LoRa message, essentially \emph{what} is leaking from \emph{how} it is leaking.  We argue, an event is something happening in the real world that can leak if it either manifests in one message (e.g. a push of a button) or a change in statistical properties of message generation (hiring/firing people at a company, production activity in industrial plant). The former causes event leakage \emph{by existence} of LoRa messages, which is something that is alarming and quite unique to this technology. The latter requires the attacker to have or build a statistical model to compare against potential anomalies (i.e. the event is characterised as rate disturbance in the LoRa message stream). This second case we address in the next section.

\subsection{Statistical Leakage (\statleak)}
With \emph{Statistical Leakage} (\statleak), we refer to cases where a \emph{deviation from normal transmission behaviour} implies the occurrence of a real-world event.
Deviation from normal behaviour can manifest in different ways in the overall traffic distribution, including, but not limited to, more (or less) traffic than expected for a short or extended period of time or a traffic inter-arrival time distribution that differs from the normal inter-arrival time. Observe that these examples of abnormal transmission behavior rely on a notion of normal or expected behaviour, however a baseline transmission behaviour is not necessary for the attacker to detect such instances, as she can detect anomalies in the transmission also without any prior knowledge~\cite{ihler2006adaptive}.

To give some examples, \statleak manifests in an industrial application every time a parking lot sensor increases its transmission profile because of more cars occupying the parking lots to attend a company event, or when a room vacancy sensor transmits less at particular times due to a series of long meetings. \statleak is not unique to LPWANs, but is particularly concerning on LPWANs because, given the specialization of the sensors, the set of events that can cause an observable deviation from the norm is very limited. Therefore, despite the fact that the traffic is encrypted, the attacker usually has a very good idea of the \emph{complete} state of the sensor.
%since sensors are usually very specialized, there are only few different conditions that could have caused and increase or decrease in traffic of a sensor.
To understand why this is important let us consider a WiFi network in which an end device suddenly starts communicating more than usual. All the attacker can say in this case is that a particular user is more active than it would otherwise have been, but there are countless servers to which the end device can be communicating with, so the attacker cannot learn much about the behaviour of a user just by observing its encrypted traffic. On the contrary, suppose we have a push button sensor communicating on a LPWAN, more traffic than usual can only mean that the button was pressed more often than usual, since to respect the LPWAN constraints the button will try to utilize the channel as little as possible, therefore the attacker gets perfect information.

An attacker exploits \statleak by observing statistical aggregates of the transmission behaviour (e.g., message counts) of a particular sensor in order to gain information about real-world events. If we assume that real-world events show up as statistical anomalies, an attacker that recognizes an anomaly in the transmission behaviour of a given sensor, over a period of time of her choosing, managed to exploit \statleak. The period of time can be as long or as short as the attacker desires. As an example, an attacker is exploiting \statleak if she can recognize that in the 30 minutes following lunch time of a particular day or that a parking lot sensor sent more packets than it would have otherwise sent in a normal day.

\subsection{Formalization}
%\aanote{`e' is used to indicate both \emph{end} and \emph{event}. Change the notations.}
We formally model an event $e$ as a tuple comprising of its start time $t_0$ and its end time $t_1$, which is $e \doteq (t_0, t_1) \in \mathbb{R}^2$ for $t_0 \leq t_1$, and we refer to the start (end) time of an event $e$ as $e.t_0$ ($e.t_1$).
Moreover, with $\mathcal{E}_{a-b} \doteq \{e_1, e_2, \dots, e_k\} \in \mathbb{R}^{2k}$ we denote a set of events that all start at or after time $t_a$ and all terminate at or before time $t_b$, that is $\forall e \in \mathcal{E}_{a-b} \Rightarrow t_a \leq e.t_0 \wedge e.t_1 \leq t_b$.

We define the trace comprising only dummy packets between times $t_a$ and $t_b$ as $\mathcal{D}_{a-b} \doteq \{d_{1}, d_{2}, \dots, d_{m}\} \in \mathbb{R}^{m}$, where $d_i$ is the time at which dummy packet $i$ was sent and $t_a \leq d_1 < d_2 < \dots < d_m \leq t_b$.
Similarly, we define a trace of real packets (i.e. non-dummy packets) sent by one ED starting from time $t_{a}$ up until time $t_{b}$ as $\mathcal{R}_{a-b} \doteq \{r_{1}, r_{2}, \dots, r_{n}\} \in \mathbb{R}^n$, where each $r_{i}$ is the time at which real packet $i$ was sent and the following condition holds: $t_a \leq r_1 < r_2 < \dots < r_n \leq t_b$.
Note that real packets are connected with events by the following relation:
\begin{equation}
r \in \mathcal{R}_{a-b} \Rightarrow \exists e \in \mathcal{E}_{a-b} | e.t_0 \leq r
\end{equation}\label{eq:ex_leak}
That is, for each message sent in the real trace, there exist a real event in the event set.
The overall transmission trace of an ED as seen by the attacker in between time $t_a$ and $t_b$ is then defined as $\mathcal{X}_{a-b} \doteq \{x_1, x_2, \dots, x_{n + m}\} \doteq \mathcal{R}_{a-b} \cup \mathcal{D}_{a-b} \in \mathbb{R}^{n+m}$, where $t_a \leq x_1 < x_2 < \dots < x_{n + m} \leq t_b$.

We model the prior knowledge of the attacker as $\pi(\mathcal{R}_{a-b})$, which is essentially the probability that a given ED will produce the real trace $\mathcal{R}_{a-b}$ in between times $t_a$ and $t_b$.
Alternatively, we can also give the attacker prior knowledge of the event distribution: $\pi_{a-b}(t_0, t_1) \doteq Pr(\exists e \in \mathcal{E}_{a-b} | t_0 \leq e.t_0 \wedge e.t_1 \leq t_1 )$, where $\pi_{a-b}(t_0, t_1)$ is essentially the probability that at least an event occurs in between time $t_0$ and $t_1$ and is defined only for $t_a \leq t_0 < t_1 \leq t_b$.

%The obfuscation mechanism is the (probabilistic or deterministic) algorithm running in the ED that decides when to send dummy messages. In our model the prior distribution of real events $\pi_{a-b}(\cdot)$ is also used by the obfuscation mechanism.
We describe the obfuscation mechanism as $q(\mathcal{X}_{a-b}| \mathcal{R}_{a-b})$, which is the probability density function (PDF) of obfuscating the real trace $\mathcal{R}_{a-b}$ with the trace $\mathcal{X}_{a-b}$. Note that, despite the fact that we give as input the obfuscated trace in the same interval as the real trace (both go from time $t_a$ to $t_b$), EDs can only place dummies in the future.

%trace Note that $q_b(\cdot)$ is only defined for dummy messages sent after time $t_b$ to reflect the fact that dummies can only be added in the future (and not in the past).
%\ipnote{Should I define the domains of all these functions?}
%\todo{add table with summary of notation}
Finally, $p(\mathcal{R}_{a-b} | \mathcal{X}_{a-b})$ is the posterior probability that the trace $\mathcal{R}_{a-b}$ of real messages was contained in the trace $\mathcal{X}_{a-b}$. Given the prior knowledge of the attacker $\pi(\mathcal{R}_{a-b})$ %which is the probability that a trace $x_{t}$ containing only real messages occurred,
and an obfuscation mechanism $q(\mathcal{X}_{a-b}| \mathcal{R}_{a-b})$, we can formally define the posterior as
\begin{equation}
	p(\mathcal{R}_{a-b} | \mathcal{X}_{a-b}) \doteq \frac{\pi(\mathcal{R}_{a-b}) \cdot q(\mathcal{X}_{a-b}| \mathcal{R}_{a-b})}{\sum_{\mathcal{R}_{a-b}' \subseteq \mathcal{X}_{a-b}} \pi(\mathcal{R}_{a-b}') \cdot q(\mathcal{X}_{a-b}| \mathcal{R}_{a-b}')}.
\end{equation}

\section{Preventing Information Leakage}
Before describing how to protect end devices (EDs) from \exleak and \statleak it is important to understand which types of events are going to leak information and which ones are not. In particular, periodic and regular transmissions such as scheduled status updates and keep alive signals do not leak information under our attacker model. This is because the attacker knows the distribution of real events and therefore observing something that she knows should be there already does not increase her information about the state of the ED. However, this does not imply that EDs that send only periodic messages will never leak any information, as the absence of a periodic transmission might also reveal information to the attacker. Observe that these corner cases of leakage can be easily obfuscated, by letting the obfuscation mechanism send a dummy message whenever a real event is not detected at the scheduled time.

Albeit trivial, this case helps in understanding one of the perfect obfuscation strategies for both \exleak and \statleak: by constantly sending messages one after the other, and filling empty transmission slots with dummies whenever there is no real information to transmit, the attacker cannot learn anything about the ED. Of course such obfuscation mechanism can never be realized in practice, because it would make the LoRa channel usable by at most one ED and the power consumption of this device would most certainly not be low, therefore not meeting the low-power requirement of devices operating on LPWANs.

Since this trivial obfuscation mechanism cannot be employed in practice, we need a way to evaluate how effective different obfuscation mechanisms are at protecting against \exleak and \statleak. Intuitively, the obfuscation mechanism needs to achieve \emph{coverage} and \emph{statistical equalization}, to protect against existential leakage and statistical leakage, respectively.
With coverage we refer to the fact that in order to make it more difficult to guess which packets are real in a given trace, the obfuscation mechanism needs to be able to cover that trace with enough likely dummy packets, i.e.\ dummy packets that are likely real in the eyes of the attacker. Statistical equalization refers instead to the ability of masking anomalies within a trace, thus equalizing or normalizing the statistics of a trace.

Towards these objectives, we introduce below the metrics that allow us to quantify how much an obfuscation mechanism, with some given transmission frequency constraints, can obfuscate a trace of real messages sent by an ED.

\subsection{Privacy Evaluation Metrics}
As previously observed in the context of location privacy~\cite{loc_privacy_otg}, a single metric usually does not simultaneously capture all the dimensions of privacy. Therefore, an obfuscation mechanism optimal for a given metric, might not be a practically desirable solution to protect the privacy of an end device. We instantiate two different metrics that can be used to complement each other when analyzing different obfuscation mechanisms: \emph{average error} and \emph{conditional entropy}.

\subsubsection{Average error}\label{sec:average_error}
This metric estimates the average error of an optimal attacker, based on a distance function between the actual real trace and the best estimation that the attacker can make based on the observation of the obfuscated trace. The distance function $d(\mathcal{R}_{a-b}, \mathcal{R}_{a-b}')$ can be defined in several ways depending on whether we are analyzing the performance of the obfuscation mechanism against \exleak or \statleak. For instance, when looking at \exleak the distance function can be defined as the absolute difference between the cardinality of $\mathcal{R}_{a-b}$ and $\mathcal{R}_{a-b}'$, while when looking at \statleak as the absolute difference between the number of statistical anomalies between $\mathcal{R}_{a-b}$ and $\mathcal{R}_{a-b}'$. The distance can also be extended to take into consideration the timing differences, and not only the absolute numbers, of the two traces.
We report here the definition of average error, $AE$:
\begin{dmath}
  AE_{a-b}(q) \doteq \int_{\mathbb{R}^{n+m}} \min\limits_{\mathcal{R}_{a-b}' \subseteq \mathcal{X}_{a-b}} \left \{ \sum\limits_{\mathcal{R}_{a-b} \subseteq \mathcal{X}_{a-b}} \pi(\mathcal{R}_{a-b}) \cdot q(\mathcal{X}_{a-b}|\mathcal{R}_{a-b}) \cdot d(\mathcal{R}_{a-b}, \mathcal{R}_{a-b}') \right \}d\mathcal{X}_{a-b}
\end{dmath}

If we consider a parking lot scenario, $AE_{a-b}(q)$ measures the average precision of an optimal attacker in guessing the number cars parking in a determinate lot on a specific interval of time, when the obfuscation mechanism $q$ is used. Similarly, by defining the distance function as a count of anomalies, $AE_{a-b}(q)$ quantifies the performance of the best attacker at identifying abnormal movements of cars in the parking lot.

\subsubsection{Conditional Entropy}
While the average error measures the error of the attacker, the conditional entropy provides a measure of how certain the attacker is, on average, that her guess is correct. As opposed to the average error, the conditional entropy does not depend on a distance function between two traces, as it is simply an information-theoretic measure on the least uncertainty an attacker can achieve. In practical terms, picking again on the parking lot scenario, say that the attacker guessed that $5$ cars parked within a given interval, the conditional entropy measures how sure the attacker can be that this event actually occurred.
%whatever the best guess of the attacker is for any observable event,
\begin{dmath}
 CE_{a-b}(q) \doteq - \int_{\mathbb{R}^{n+m}}\sum\limits_{\mathcal{R}_{a-b} \subseteq \mathcal{X}_{a-b}}\pi(\mathcal{R}_{a-b}) \cdot q(\mathcal{X}_{a-b}|\mathcal{R}_{a-b})\cdot\log(p(\mathcal{R}_{a-b} | \mathcal{X}_{a-b}))d\mathcal{X}_{a-b}
\end{dmath}
%This metric can also be viewed in terms of how many bits of information are necessary on average to identify every real message in the obfuscated trace.

\subsection{Are \exleak and \statleak fundamentally the same?}
Leakage by existence refers to scenarios where observation of a single message can leak the occurrence of a certain real-world event. In statistical leakage, we summarize scenarios where statistical properties of aggregates of messages leak the occurrence of events, which reveal themselves as anomalies of the message count. More precisely, an anomaly is given by a message rate disturbance, observed at a certain time scale (which depends on the nature of the event). The question arises, whether we can conceptually reduce \statleak to \exleak by applying a transformation that replaces anomalies with individual messages and ignores all remaining transmissions. Without obfuscation, this equivalence hypothesis indeed holds.
%\todo{describe better the transformation}

However, obfuscation changes this. A fundamental property of leakage by existence is that the detailed time information of the event cannot be completely removed by an obfuscation strategy that only involves dummies. However, statistical anomalies can potentially be hidden completely. The reason for this is that statistical leakage inherently assumes a discretization on the part of the attacker, thereby allowing an obfuscator to potentially max out the conditional entropy (e.g. by waterfilling the rate subject to the attacker's discretization interval). This discretization inherently has the property that it happens over time intervals that are likely to involve multiple non-dummy transmissions. Otherwise, there would be nothing to aggregate. This means, that it is - at least from the perspective of instantaneous channel occupancy - possible to maintain this rate by adding dummies. This denotes the fundamental difference between leakage by existence and statistical leakage. In leakage by existence, the fundamental time discretization is given by the application's time resolution. One could argue that there still is a fundamental limit on time information. However, waterfilling at the application time-resolution is fundamentally out of reach for any obfuscator, due to limits in channel access rates as well as our focus on delay-intolerant applications. Therefore, complete event hiding is out of reach in leakage by existence.
%\ipnote{The message here is that you can equalize rate, with a reasonable number of messages (I guess given some assumptions on the max transmission rate), but you cannot do the same for single messages because as discussed before this is too many messages (essentially send back to back).}

In both \exleak and \statleak, an attacker performs for a certain point in time or time interval a binary hypothesis test on a certain \emph{observable}, mapping it to an event or not. In leakage by existence, the event observables are given by the messages. In statistical leakage, not all intervals containing events qualify as event observable (they have to register as anomaly at the attacker). There, the attacker performs some kind of aggregation over an interval, the output of which is some real number, on which a binary decision is applied. This number, let's call it \emph{anomaly indicator}, is affected by both real and dummy messages. Dummy messages might reduce the anomaly indicator at certain intervals and increase it at others, both removing observable events and adding them (as opposed to \exleak, where observables can only be added).

\begin{figure*}[t]
	\begin{center}
		\includegraphics[scale=0.9]{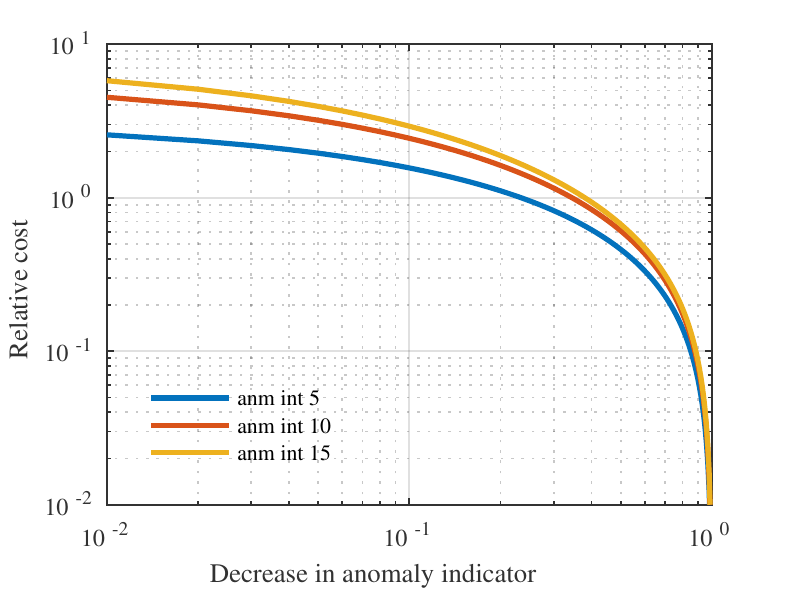}
		\includegraphics[scale=0.9]{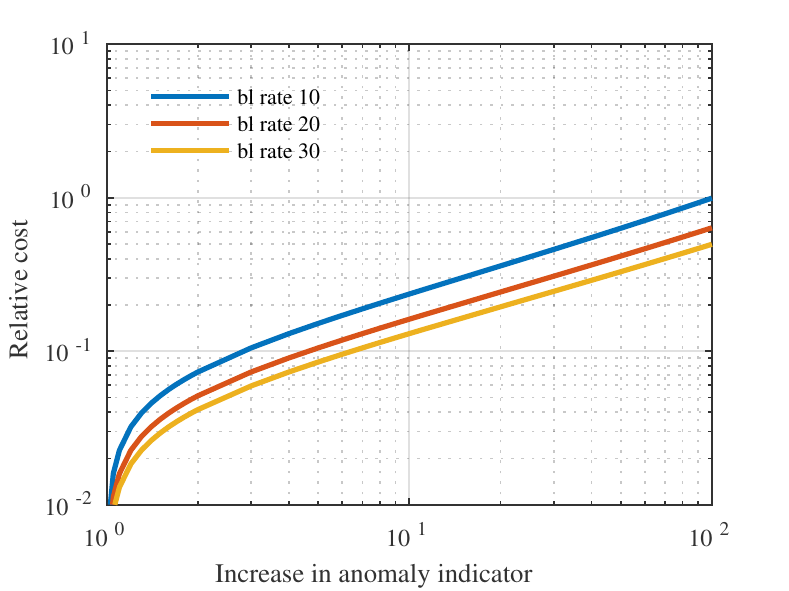}
		\caption{Obfuscation cost associated with influencing the expected anomaly indicator (index of dispersion). The relative cost for increasing the anomaly indicator depend on the baseline (bl) rate whereas the cost for reduction depends on the relative intensity of the anomaly (anm int).}
		\label{fig:cost_wf_and_fake}
	\end{center}
\end{figure*}

\subsection{Building blocks of an obfuscator}

In the following, we address the most important practical concerns and challenges faced in the design of an obfuscation mechanism against both statistical and existential leakage.

\paragraph{Background traffic model} An obfuscator needs to maintain a time-dependent model for the background traffic. Traffic generated by human activity exhibits periodicity at different time scales which may itself not contain meaningful event information to hide and therefore needs to be separated. Such contributions have previously been modelled as multiplicative components of a time-dependent poisson rate in the context of event detection~\cite{ihler2006adaptive}.

\paragraph{Event prior} An obfuscator requires some prior knowledge about the occurrence of events worth obfuscating for any adaptive mechanism. For statistical leakage, this is given by a time scale at which an event occurs (aggregation interval) as well as at which time it occurs. In order to avoid existential leakage, an obfuscator requires knowledge about the temporal distribution of individual messages.

\paragraph{Attacker prior/intent} Complete information hiding against statistical leakage can only be provided by constant high-rate transmissions. This is not practical. Therefore, a meaningful obfuscation mechanism needs to have a certain attacker in mind. An attacker's temporal prior is partly given by the overall observation interval as well as the aggregation interval.

\paragraph{Multiresolution obfuscation} In general, an obfuscator has to provide protection against statistical leakage at multiple time scales. Obfuscation mechanisms at different time scales could also interact in an unfavourable way, e.g. lower scale obfuscation by waterfilling could amplify a higher scale anomaly.

\paragraph{Temporal dummy placement} Excess messages placed for statistical leakage protection need to pass the same logic as dummies for existential leakage protection. This means the temporal placement of any dummy message has to blend in with the temporal distribution of real messages. Otherwise, dummy traffic for avoiding statistical leakage could be easily separated based on more fine-grained temporal statistics.

%!TEX root =  lora-privacy-main.tex
\section{Evaluation}

In the following, we attempt to shed light on the feasibility  of obfuscation against statistical leakage under a power constraint.

\subsection{Reducing the problem to a single timescale}

Fundamentally, with statistical leakage we refer to event information that is leaked by an attacker observing the statistics of message counts. This is not limited to a certain timescale. In this evaluation however we simplify the problem to binary information that leaks over a single timescale. We therefore assume time to be discretized in intervals of constant length $S$. We model the background (baseline) traffic to be statically distributed with poisson rate $\lambda$, contributing with $S$ independent samples to the overall process $\lambda S$.
\footnote{While this might at first seem unrealistically simple, real event traces have actually been modelled with time-varying poisson distributions~\cite{ihler2006adaptive}. It is also worth noting that many inhomogeneities are periodic and can potentially be removed by an attacker (e.g., day-night patterns). The static poisson distribution captures the remainder in the count distribution that is purely random, reflecting an attacker that is well-informed about systematic changes in the distribution.}
We model event information as a singular variation of the rate in one of the slots within $S$. An interval of length $S$ that contains a slot with poisson rate $\lambda_A > \lambda$ is considered an \emph{anomaly}. We define the \emph{anomaly intensity} $I = \frac{\lambda_A}{\lambda}$ as a measure for the magnitude of the anomaly. We consider each interval to contain an anomaly with an equal probability $R_p$ (positive rate).

An attacker's goal is to detect such anomalies. For this purpose, we consider the attacker to rely on a single observable per interval, the index of dispersion $D=\sigma^2/\mu$, where $\sigma^2$ refers to the variance of the message count observed in the $S$ slots divided by the respective mean $\mu$. The index of dispersion is suited in order to distinguish random (poisson) from bursty event traces~\cite{cox1955some}. The attacker performs for each interval a binary hypothesis test on the index of dispersion
\[\mathcal{H}_0: D = 1\]
\[\mathcal{H}_1: D > 1,\]
where $\mathcal{H}_0$ denotes the null hypothesis (no anomaly) and $\mathcal{H}_1$ the alternative hypothesis (anomaly). We assume the attacker knows the distribution of $D$ under both hypotheses.

An obfuscator can fundamentally try to hide real anomalies by waterfilling as well as introduce fake anomalies. An obfuscation strategy is given by the probabilities for waterfilling given an anomaly ($P_{wf}$) as well as the probability for adding a fake anomaly given no anomaly ($P_{f}$). Both measures are associated with a cost in terms of power.

\subsection{Cost of \statleak obfuscation}

Power is an important constraint in any LPWAN application. Therefore, we model the obfuscation strategy to be informed by the relative power cost per time interval. Since messages and dummies need to be indistinguishable anyway for effective obfuscation we consider the relative cost in terms of relative message counts of dummies vs. real traffic.

\subsubsection{Adding fake anomalies}
First, we consider the case where the obfuscator transmits dummy messages within an interval that consists only of background traffic. This corresponds to adding an excess rate of $\lambda_{fa}$ in one of the $S$ slots of the interval. The mean becomes
\[\mu' = \frac{(S-1)\lambda+(\lambda + \lambda_{fa})}{S},\]
the variance
\[{\sigma'}^2 = \left[\frac{(S-1)\lambda^2 - (\lambda + \lambda_{fa})^2}{S} - {\mu'}^2\right]\frac{S}{S-1} + \mu'.\]
The expected cost associated with increasing the index of dispersion by a factor of $k$ can be found analytically by solving $D' = kD = {\sigma'}^2/\mu'$ for $\lambda_{fa}$.
%The right hand side plot in Figure~\ref{fig:cost_wf_and_fake} shows this value for different background rates.
As only one slot is affected, the cost for adding a fake anomaly is given by the fake anomaly excess rate itself. The relative cost is given by
\[C_{f} = \frac{\lambda_{fa}}{\lambda S}\]

\subsubsection{Waterfilling}
Similarly, by waterfilling the obfuscator changes the statistics of the traffic by lifting the overall mean over the interval to
\[\mu' = \frac{(S-1)(\lambda + \lambda_{wf}) + \lambda I}{S},\]
as well as the variance to
\[{\sigma'}^2 = \left[\frac{(S-1)(\lambda + \lambda_{wf})^2 - (\lambda I)^2}{S} - {\mu'}^2\right]\frac{S}{S-1} + \mu'.\]
Again, the expected cost of decreasing an anomaly by a factor of $k$ can be found by solving $D' = D/k = {\sigma'}^2/\mu'$ for $(S-1)\lambda_{wf}$. This value is shown for some anomaly configurations in the left side plot of Figure~\ref{fig:cost_wf_and_fake}.
The expected relative waterfilling cost for the entire interval amounts to
\[C_{wf} = \frac{\lambda_{wf}(S-1)}{\lambda S + \lambda_A}.\]

In Figure~\ref{fig:cost_wf_and_fake} we show the dependency of the relative change on the anomaly indicator (index of dispersion) for both obfuscation mechanisms. Note that waterfilling leads to a decrease in the anomaly indicator, while a fake anomaly is associated with an increase. Overall, a fake anomaly is less costly than waterfilling for a similar relative shift of the attacker observable.

\subsection{Obfuscation Strategy}

After introducing the attacker and formulating the specific power constraint considered, we will establish the conditions for both optimal and sub-optimal obfuscation and motivate the behaviour of the obfuscator under our model.

\subsubsection{Attacker knowledge}
We assume the attacker to know the statistics of the observable (index of dispersion) under both hypotheses. Moreover, the attacker knows the relative rate of anomalies (i.e. rate of positives) $R_p$. Correspondingly, since we are dealing with a binary detection problem, the rate of negatives (no anomaly) is given by $R_n = 1-R_p$. The attacker also shares the same time discretization into intervals of equal length S with the obfuscator. The width of $S$ can be considered a temporal prior, i.e. the attacker has some knowledge about the temporal duration of the anomalies he wants to detect. We assume the attacker not to include the overall message count of intervals into his decision. The reasoning behind this is as follows. An attacker that includes changes in the overall rate over time essentially performs anomaly detection at a higher timescale. While this does not exclude the possibility of statistical leakage of higher-scale anomalies, the same obfuscation mechanisms could also be applied at this higher scale, at the cost of a corresponding relative cost.

\subsubsection{Power constraint}
As introduced, waterfilling and fake anomalies differ in terms of their relative costs, $C_{wf}$ and $C_{f}$. $P_{wf}$ and $P_f$ are the probabilities that the obfuscator employs waterfilling or places a fake anomaly, respectively, and together denote the obfuscation strategy. We consider a fixed power budget for the dummy generation in order to account for LoRa low-power goal. Specifically, the maximum expected relative obfuscation cost cannot exceed the overall background message rate:

\begin{equation}
R_{p}P_{wf}C_{wf} + (1-R_{p})P_{f}C_{f} \stackrel{!}{\leq} 1
\label{eq:power_constraint}
\end{equation}

\subsubsection{Complete obfuscator knowledge}
%We first consider an optimal strategy for limiting statistical information leakage in our simplified scenario.
First, we assume the obfsucator to have optimal knowledge w.r.t. the time intervals in which anomalies occur. The obfuscator's strategy is defined by his choice for the probability at which anomalies are waterfilled $P_{wf}$ as well as the probability at which fake anomalies are added to intervals without anomalies, $P_{f}$. Under optimal \statleak obfuscation, the observation of the index of dispersion subject to a certain scale (i.e. discretization into intervals of length S) should not help the attacker in deciding which intervals contain anomalies.
%I.e. the probability of observing an anomaly should be independent of the observable (the index of dispersion).

We introduce a parameter $\varepsilon$, capturing the degree of sub-optimality of obfuscation, and define the obfuscator's goal subject to this parameter as
\begin{equation}
\frac{R_{p}P_{wf}}{R_{n}(1-P_{f}) + R_p P_{wf} } \cdot (\varepsilon + 1) \stackrel{!}{=} \frac{R_p(1-P_{wf})}{R_p(1-P_{wf})+R_n P_{f}}.
\label{eq:optimal_obf_complete}
\end{equation}
Since $R_n= 1-R_p$, the above equation holds true for optimality of obfuscation (i.e. at $\varepsilon=0$), whenever
\[P_{wf}=1-P_{f}.\]

\subsubsection{Incomplete obfuscator knowledge}
Until now, we assumed an obfuscator with complete knowledge about the times at which anomalies occur. However, this might be an unrealistic assumption. Therefore, in addition to a scenario with full knowledge, we model an obfuscator with limited knowledge. This limitation is reflected in a likelihood $P_{tp}<1$ of predicting an anomaly and $P_{tn}<1$ of correctly anticipating baseline traffic. We assume the obfuscator to be aware of this uncertainty. The condition for $\varepsilon+1$-obfuscation, originally formulated in Equation \ref{eq:optimal_obf_complete}, becomes
\begin{equation}
\frac{R_{p}P_{tp}P_{wf}}{R_{n}(1-P_{tn}P_{f}) + R_p P_{tp}P_{wf} } \cdot (\varepsilon + 1) \stackrel{!}{=} \frac{R_p(1-P_{tp}P_{wf})}{R_p(1-P_{tp}P_{wf})+R_n P_{tn}P_{f}},
\label{eq:optimal_obf_incomplete}
\end{equation}
which is under assumption of optimality ($\varepsilon = 0$) satisfied for the obfuscation parameters
\[P_{wf} = \frac{1-P_{tn}P_{f}}{P_{tp}}\]
and
\[P_{f} = \frac{1-P_{tp}P_{wf}}{P_{tn}}.\]

\subsubsection{Sub-optimal obfuscation}
Optimal obfuscation cannot be achieved whenever Equations~\ref{eq:optimal_obf_complete}~or~\ref{eq:optimal_obf_incomplete} cannot be satisfied with $\varepsilon = 0$ and $P_{f},P_{wf}\in [0, 1]$, under the power constraint. In that case, we assume the obfuscator to choose the obfuscation strategy, i.e. $P_{f}$ and $P_{wf}$, such that $\vert\varepsilon \vert$, i.e. the relative bias in the posterior probability distribution s.t. both observations, is minimized.

\subsection{Evaluation results}
In the evaluation, we explore the following two directions. First, we characterize the parameter space, aiming to understand which combinations of values for the rate of anomalies $R_P$ and the anomaly intensity $I$ allow for optimal obfuscation. Second, we would like to quantify the information leakage under sub-optimal obfuscation, using the metrics introduced in the previous section.

\subsubsection{Optimal obfuscation}
In this evaluation, we aim to understand the parameter space of an optimal obfuscator. That is, which combinations of anomaly rate $R_P$ and anomaly intensity $I$ allow us to build an optimal obfuscator under some power constraint. Specifically, we are interested to know which parameter combinations satisfy Equation~\ref{eq:power_constraint}, given the dependency of the obfuscation cost on the anomaly intensity $I$.
%Figures \ref{fig:metrics_complete_kn} and \ref{fig:metrics_incomplete_kn} display the obfuscator strategy parameters fulfilling Equation \ref{eq:power_constraint} with equality.
Considering Figures \ref{fig:metrics_complete_kn} and \ref{fig:metrics_incomplete_kn}, obfuscation is optimal where the metric coincides with the ideal value. This means that, at anomaly rates where this is the case, values for $P_f$ and $P_{wf}$ exist that satisfy the power constraint.
%The relationship between $P_{wf}$ and $P_{f}$ is thereby defined by the the obfuscator's uncertainty.
We can see in both scenarios that increasing the anomaly intensity decreases the width of the anomaly rate intervals for which an optimal strategy exists. Moreover, the support of the optimal strategy over the anomaly rate also decreases for all anomaly intensities when the knowledge of the obfuscator decreases. In the second scenario, we considered an obfuscator which correctly guesses $99\%$ of baseline intervals and $70\%$ of anomaly intervals. This uncertainty of the obfuscator makes it impossible to make optimal choices for anomaly rates under $50\%$, given an anomaly intensity that is $30$ or higher. However, we argue that it is anomalies at these low occurrence rates deserving the name in the first place, indicating rare events that might be important to hide.
We hence conclude that, if an obfuscator is limited in knowledge, perfect hiding of statistical leakage is not possible at relative anomaly intensities above $30$, given the constraint that power spending for obfuscation may not exceed the power used for actual traffic.

\subsubsection{Sub-optimal obfuscation}
\begin{figure*}[t]
	\begin{center}
		\includegraphics[scale=0.9]{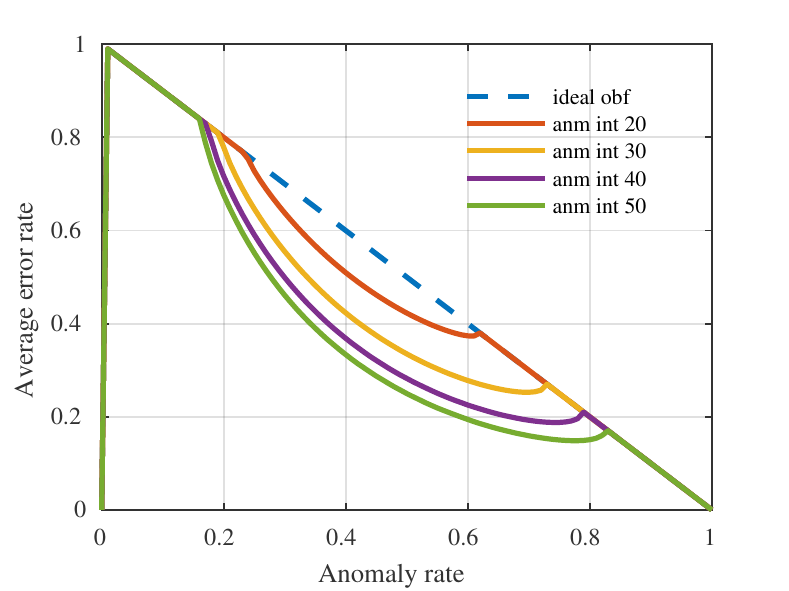}
		\includegraphics[scale=0.9]{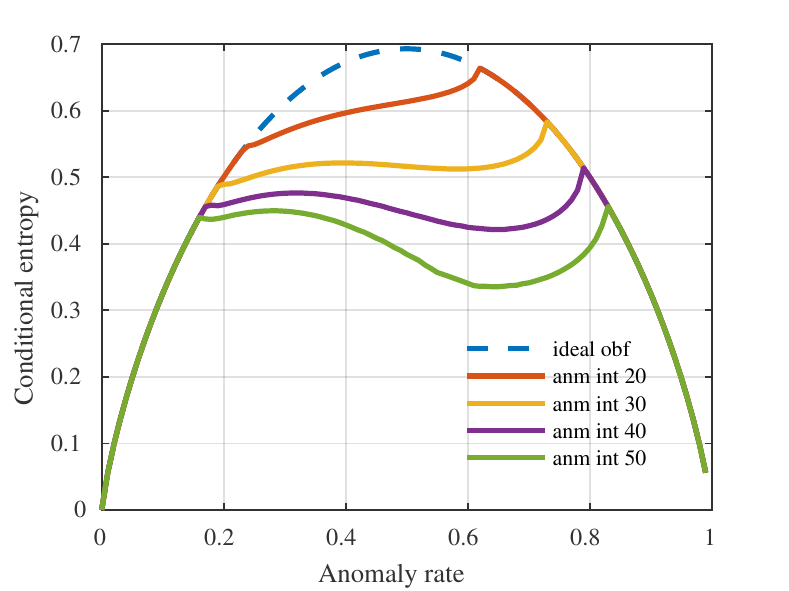}
		\caption{For different anomaly intensities, we show the attacker's error rate in guessing anomalies as well as the conditional entropy. We also highlight the theoretical value of the metrics for optimal obfuscation. Where the results differ from this curve, optimal obfuscation cannot be provided under an equivalent power constraint of 1. In the scenario shown, we assume the obfuscator to have optimal knowledge about anomaly occurences.}
		\label{fig:metrics_complete_kn}
	\end{center}
\end{figure*}

\begin{figure*}[t]
	\begin{center}
		\includegraphics[scale=0.9]{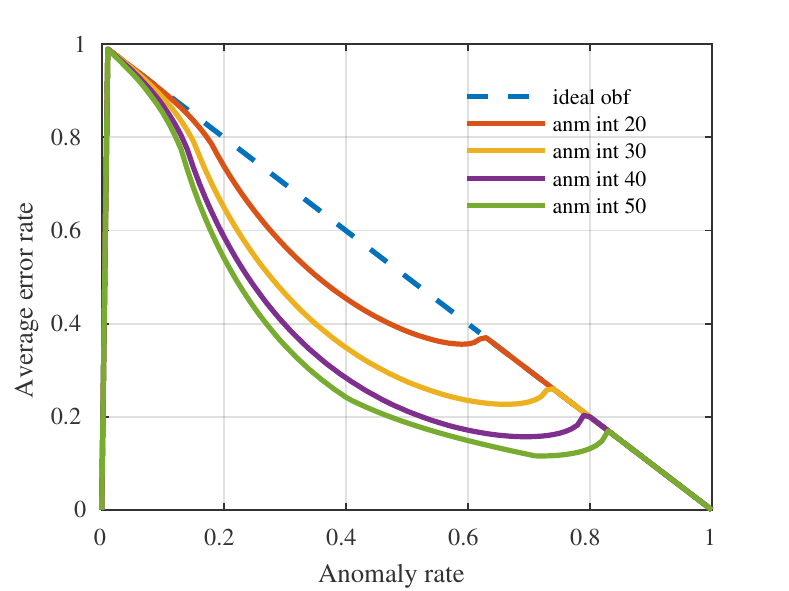}
		\includegraphics[scale=0.9]{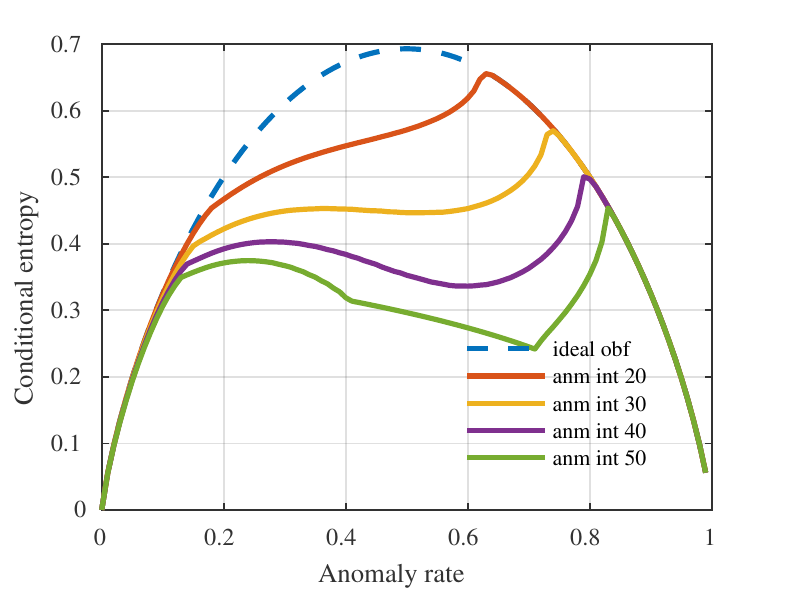}
		\caption{If we limit the knowledge of the obfuscator 99$\%$ (TNR) and 70$\%$ (TPR) certainty for baseline and anomaly, respectively, the domain of anomaly rates that can be ideally protected under an equal power constraint of 1 shrinks considerably. In general, we observe a decrease in both the attacker's guessing error as well as the conditional entropy compared to obfuscation with full knowledge about anomaly occurences.}
		\label{fig:metrics_incomplete_kn}
	\end{center}
\end{figure*}

In Figures \ref{fig:metrics_complete_kn} and \ref{fig:metrics_incomplete_kn}, deviations from the ideal value characterize information leakage under sub-optimal obfuscation. The ideal value refers to a guessing attacker that only uses his prior knowledge about $R_P$, and therefore does not learn anything by observing the trace. We evaluated guessing error and conditional entropy for both complete and incomplete obfuscator knowledge. The attacker is modelled to guess anomalies and baseline intervals with the respective probabilities at which they occur. The guessing error is defined as the ratio of anomalies that is expected to be not detected by the attacker. Consistently with the previous result, we can observe that the range of optimal obfuscation shrinks for increasing anomaly intensities. Moroever, depending on the anomaly rate the attacker's error rate decreases, especially for more pronounced anomalies. For example, at a 20$\%$ rate of anomalies with intensity 40, the power constraint results in a 20$\%$ decrease in anomaly guessing error for the attacker.

\subsection{Multiple timescales}
We note that, despite the evaluation being confined to a single timescale, the findings are of importance also for applications that potentially leak at different timescales. Statistical patterns of real applications can convey event information at multiple, different timescales~\cite{ihler2006adaptive}. Hiding anomalies at one timescale hence only constitutes a necessary condition for overall obfuscation. Moreover, assuming that anomalies at different timescales occure independently, we can expect the overall relative cost to be given by a product of contributions at different scales. In particular, by providing protection against existential leakage at all times, an obfuscator would be forced to place dummies in the vicinity of every real message, thereby multiplying the background rate $\lambda$ by the size of the anonymity set provided.

Our results for one scale indicate that protection against statistical leakage at one scale is bound to collide with fundamental power limitations in the LPWAN context, a problem that is expected to get even worse if more scales and existential leakage are also considered by an obfuscator.

%!TEX root =  lora-privacy-main.tex
\section{Related Work}
The problem of preventing information leakage in modern communication and networking technologies has received much attention due to the rapid growth of applications that rely on these technologies to exchange sensitive data. Privacy issues in wireless sensor networks have been studied mainly with the goal of protecting the network's spatial~\cite{shao2008towards} and temporal~\cite{kamat2007temporal} information. Many works~\cite{kamat2005enhancing, mehta2007location, shao2008towards, xi2006preserving, yang2008towards} have studied the problem of protecting source location in wireless sensor networks, i.e. where a particular event originated.
The proposed solutions leverage the multi-hop topology of the network to generate phantom~\cite{kamat2005enhancing} and random routes~\cite{xi2006preserving} with fake data to achieve one static spatial distribution. Due to the delay-tolerant nature of the applications, many solutions to protect temporal privacy in wireless sensor networks involve delaying or aggregating messages~\cite{kamat2009temporal}. The above solutions do not work in the context of LPWANs, as many of their applications are delay-intolerant and rely on event-triggered communications for their operation. Furthermore, the one-hop network topology of LPWANs makes the location-hiding schemes proposed for wireless sensor networks infeasible.

In the context of general networks, information leakage has been extensively studied through a variety of anonymity and mix network mechanisms. The main goal of mix networks is to hide the association between the sender and the receiver. Several attacks~\cite{syverson2001towards, back2001traffic, back2001freedom, serjantov2003passive} have leveraged the timing information of mix network packets to derive a relationship between senders and receivers. The proposed countermeasures involve packet dropping~\cite{levine2004timing}, constant and adaptive dummy injections~\cite{shmatikov2006timing}, introducing artificial packet delays~\cite{kamat2009temporal}, and so on. These countermeasures were designed for general networks where the participating nodes do not have any power constraints. On the other hand, LPWANs were designed specifically to enable ultra low-power operation of the end devices.

The recent works on LPWANs and LoRa have focussed on addressing challenges related to network performance, coverage and scalability~\cite{bor2016lora,georgiou2017low,adelantado2017understanding,magrin2017performance}. Several vulnerabilities, such as replay attacks~\cite{tomasin2017security,aras2017exploring}, reactive jamming~\cite{aras2017selective}, key extraction~\cite{tomasin2017security} and fingerprinting end devices~\cite{Robyns:2017:PFL:3098243.3098267} were successfully demonstrated. To the best of our knowledge, the implications of information leakage in LPWANs have so far not been studied. This paper therefore provides a first insight on the privacy guarantees of LPWANs.

\section{Conclusion}
In this work, we showed that the event-driven communication strategy adopted by LPWANs to save power inherently leaks information, independently of the implemented cryptographic primitives. Furthermore, we demonstrated that it is hard to implement privacy enhancing techniques on LPWANs without incurring significant communication and computational (energy) cost. Given the wide variety of privacy sensitive applications that are beginning to rely on LoRa and other LPWAN technologies, we have highlighted through this paper the challenges that exist in designing low-power, low-cost, wide area networks and guaranteeing strong privacy.

\section{Acknowledgements}
This project has received funding from the European Research Council (ERC) under the European Union’s Horizon 2020 research and innovation programme under grant agreement N\textsuperscript{o} 726227.

\bibliographystyle{ACM-Reference-Format}
\bibliography{lora}

\end{document}